\documentclass[
reprint,
superscriptaddress,
 amsmath,amssymb,
 aps,
]{revtex4-2}

\usepackage[caption=false]{subfig} % APS-compatible subfigures
\usepackage{graphicx}% Include figure files
\usepackage{dcolumn}% Align table columns on decimal point
\usepackage{bm}% bold math
\usepackage{mathtools} % for \mathclap
\usepackage{hyperref}% add hypertext capabilities
\usepackage{graphicx,color,hyperref,xcolor}
\usepackage{braket} % for \ket and related commands

\begin{document}

\title{Engineering Vacuum Fluctuations in Hyperbolic Heterostructures}

\author{Jie-Cheng Feng}
\email{jiecheng.feng@mpsd.mpg.de}
\affiliation{Max Planck Institute for the Structure and Dynamics of Matter, Center for Free Electron Laser Science, Luruper Chaussee 149, 22761 Hamburg, Germany}
\affiliation{Institut für Theorie der Statistischen Physik, RWTH Aachen University, 52056 Aachen, Germany}

\author{Xinle Cheng}
\affiliation{Max Planck Institute for the Structure and Dynamics of Matter, Center for Free Electron Laser Science, Luruper Chaussee 149, 22761 Hamburg, Germany}

\author{Itai Keren}
\affiliation{Department of Physics, Columbia University, New York, New York 10027, USA}

\author{Dante M. Kennes}
\affiliation{Max Planck Institute for the Structure and Dynamics of Matter, Center for Free Electron Laser Science, Luruper Chaussee 149, 22761 Hamburg, Germany}
\affiliation{Institut für Theorie der Statistischen Physik, RWTH Aachen University, 52056 Aachen, Germany}

\author{Abhay N. Pasupathy}
\affiliation{Department of Physics, Columbia University, New York, New York 10027, USA}
\affiliation{Condensed Matter Physics and Materials Science Division, Brookhaven National Laboratory, Upton, NY, USA}

\author{Dmitri N. Basov}
\affiliation{Department of Physics, Columbia University, New York, New York 10027, USA}

\author{Emil Viñas Boström}
\email{emil.bostrom@mpsd.mpg.de}
\affiliation{Max Planck Institute for the Structure and Dynamics of Matter, Center for Free Electron Laser Science, Luruper Chaussee 149, 22761 Hamburg, Germany}

\author{Angel Rubio}
\email{angel.rubio@mpsd.mpg.de}
\affiliation{Max Planck Institute for the Structure and Dynamics of Matter, Center for Free Electron Laser Science, Luruper Chaussee 149, 22761 Hamburg, Germany}
\affiliation{Nano-Bio Spectroscopy Group, Departamento de Física de Materiales, Universidad del País Vasco, 20018 San Sebastian, Spain}
\affiliation{Initiative for Computational Catalysis (ICC), Flatiron Institute, 162 Fifth Avenue, New York, New York 10010, USA}

\date{\today}

\begin{abstract}
Vacuum fluctuations can be engineered using optical cavities, resonators, and surface-polaritonic modes, but extending such control deep inside a material remains challenging.
Here we show that the vacuum field inside a hyperbolic material can be reshaped by a much thinner overlayer with opposite hyperbolicity.
This effect requires overlap between the hyperbolic bands of the two materials and modifies the coupling between the vacuum field and matter excitations within the bottom material, without changing its intrinsic dielectric properties.
As an application, we use Eliashberg theory to study a two-dimensional superconducting layer inside the bottom material, showing that the overlayer can remotely change its critical temperature.
Our results may shed light on recent experiments reporting changes in superconductivity arising from a frequency-matched hyperbolic overlayer~\cite{kerenCavityalteredSuperconductivity2026a}, and suggest an on-chip route to long-range control of matter excitations and collective material properties.
\end{abstract}

\maketitle

Vacuum fluctuations are an unavoidable feature of the quantized electromagnetic field. 
Although not directly observable, they underlie many fundamental phenomena, such as the Purcell effect~\cite{ProceedingsAmericanPhysical1946,goyObservationCavityEnhancedSingleAtom1983}, the Lamb shift \cite{lambFineStructureHydrogen1947,betheElectromagneticShiftEnergy1947}, and the Casimir force~\cite{casimir1948attraction,sparnaay1958measurements}. 
Recent work further suggests that interactions with the electromagnetic vacuum may influence both material properties~\cite{appuglieseBreakdownTopologicalProtection2022,jarcCavitymediatedThermalControl2023,enknerTunableVacuumfieldControl2025,kerenCavityalteredSuperconductivity2026a,xuVacuumdressedSuperconductivityNbN2026,montanaroCavityenhancedSuperconductingResponse2026,zhangCavityEnhancedSuperconductivity2026,wangEvidenceVacuumenhancedSuperconductivity2026} and chemical reaction rates~\cite{thomasGroundStateChemicalReactivity2016,thomasTiltingGroundstateReactivity2019,ahnModificationGroundstateChemical2023}. 
A common route to realize such effects is to place matter in a cavity or resonator, tailored to confine the field and enhance the electromagnetic mode density and effective light--matter coupling~\cite{sentefCavityQuantumelectrodynamicalPolaritonically2018d,hagenmullerEnhancementElectronPhonon2019,hubenerEngineeringQuantumMaterials2021a,garcia-vidalManipulatingMatterStrong2021a,schlawinCavityQuantumMaterials2022a,luCavityEngineeringSolidstate2025,bretscherFluctuationEngineeringCavity2026}. 
However, in this approach the vacuum field is typically considered as a fixed property of the environment surrounding the device, rather than a dynamical and controllable quantity.

Here we explore how to engineer not only the magnitude of the vacuum fluctuations, but also their directionality and the extent over which they are modified inside a material.
Such control promotes the electromagnetic vacuum from a passive background to an active design element of the material platform. Hyperbolic materials (HMs), which are anisotropic media in which different principal components of the dielectric tensor have opposite signs~\cite{poddubnyHyperbolicMetamaterials2013,daiTunablePhononPolaritons2014a,caldwellSubdiffractionalVolumeconfinedPolaritons2014,guoHyperbolicMetamaterialsDispersion2020,leeHyperbolicMetamaterialsFusing2022}, provide a natural setting for this idea. The hyperbolic dispersion allows large-momentum modes to propagate inside the material while remaining evanescent in free space, leading to light propagation beyond the diffraction limit and to a large photonic density of states~\cite{smithElectromagneticWavePropagation2003,podolskiyStronglyAnisotropicWaveguide2005,jacobEngineeringPhotonicDensity2010}.
This physics was first explored extensively in artificial hyperbolic metamaterials, where it was used to realize phenomena such as hyperlensing, negative refraction, and enhanced spontaneous emission~\cite{liuFarFieldOpticalHyperlens2007,hoffmanNegativeRefractionSemiconductor2007,krishnamoorthyTopologicalTransitionsMetamaterials2012}.
More recently, natural van der Waals (vdW) HMs, especially hexagonal boron nitride (hBN), have provided a low-loss platform for infrared polaritonic modes with deep-subwavelength confinement~\cite{daiTunablePhononPolaritons2014a,caldwellSubdiffractionalVolumeconfinedPolaritons2014,yoxallDirectObservationUltraslow2015,daiSubdiffractionalFocusingGuiding2015a,narimanovNaturallyHyperbolic2015,gilesUltralowlossPolaritonsIsotopically2018,caldwellPhotonicsHexagonalBoron2019,herzigsheinfuxHighqualityNanocavitiesMultimodal2024a}.
They have therefore become a versatile platform for quantum optics in two-dimensional materials, including strong and ultrastrong light--matter coupling~\cite{ashidaCavityQuantumElectrodynamics2023,andolinaQuantumElectrodynamicsGraphene2026}, light emission~\cite{guoHyperbolicPhononpolaritonElectroluminescence2025}, long-range transport of energy or information~\cite{cortesSuperCoulombicAtomAtom2017a,huTransportPolaritonsHyperbolic2026,alvarez-perezLongRangeMidInfraredEnergy2026,fengColorCentersHyperbolic2026}, and the control of material properties~\cite{kerenCavityalteredSuperconductivity2026a,dongEffectHyperbolicPhoton2026}.
In the following, we focus specifically on hyperbolic phonon-polariton systems, where the electromagnetic field is tied to anisotropic ionic polarization, while the same calculations for electronic, plasmonic, or excitonic hyperbolic materials are expected to lead to similar conclusions.

\begin{figure}[hbt]
    \centering
    \includegraphics[width=0.23\textwidth]{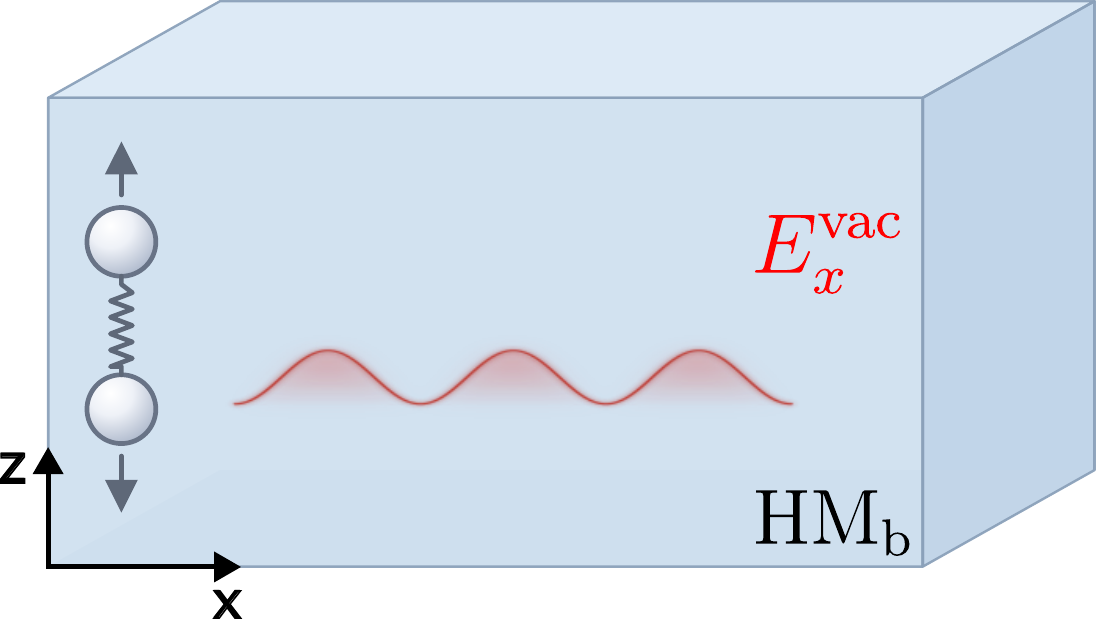}
    \hfill
    \includegraphics[width=0.24\textwidth]{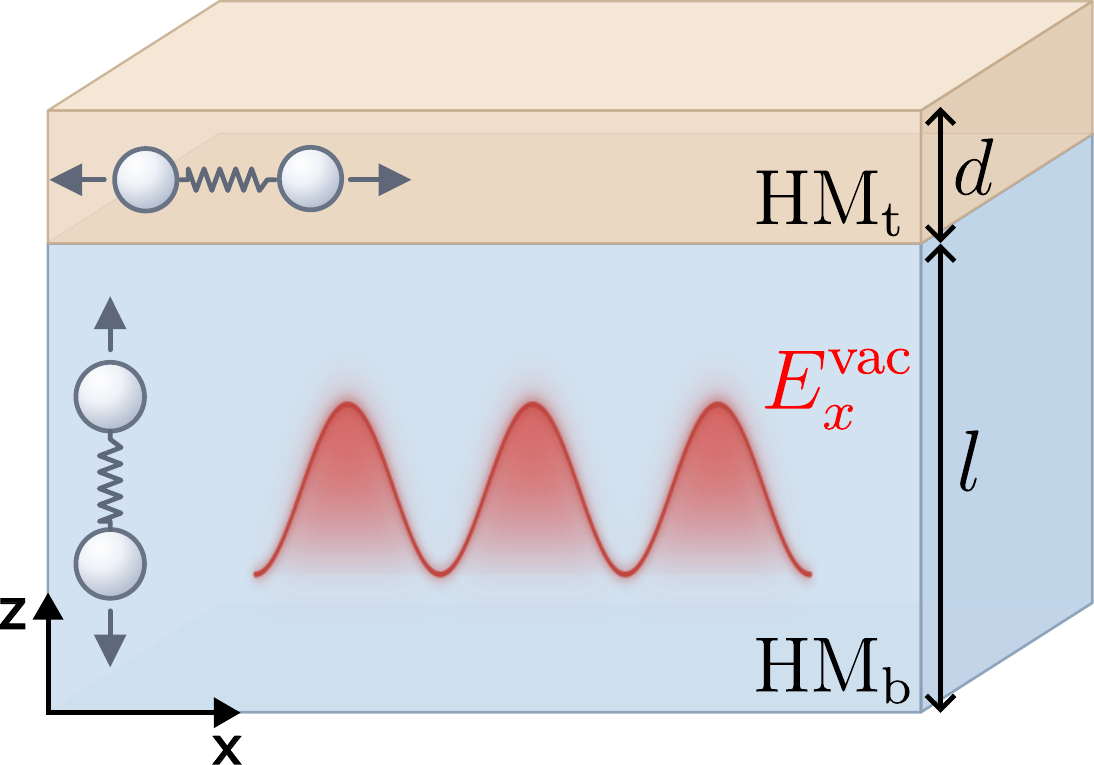}
    \caption{
    {\bf Directional control of vacuum fluctuations in a hyperbolic phonon polariton heterostructure.}
    Left: In a bare hyperbolic material $\mathrm{HM}_\text{b}$, the relevant phonon-polariton mode is dominated by a material excitation with out-of-plane polarization, such that the vacuum fluctuations of the in-plane electric field $\langle E_x^2\rangle_{\rm vac}$ are weak.
    Right: $\mathrm{HM}_\text{b}$ is capped by a second hyperbolic material $\mathrm{HM}_\text{t}$, with differently oriented anisotropy arising from a material excitation with primarily in-plane polarization.
    When the hyperbolic frequency windows overlap, the polariton modes hybridize and enhance $\langle E_x^2\rangle_{\rm vac}$ deep inside $\mathrm{HM}_\text{b}$.}
    \label{fig:setup}
\end{figure}

In this Letter, we show that adding a top hyperbolic layer can reshape the vacuum field and its coupling to matter excitations throughout a bottom hyperbolic material, without changing the bottom material’s intrinsic dielectric properties.
We obtain this spatial and directional control using two hyperbolic materials with orthogonally oriented anisotropy, stacked as illustrated in Fig.~\ref{fig:setup}.
When the two hyperbolic frequency windows overlap, which in hyperbolic phonon-polariton systems correspond to the LO--TO Reststrahlen bands, polaritons in the two materials hybridize across the interface~\cite{sternbachNegativeRefractionHyperbolic2023,berkowitzHyperbolicCooperPairPolaritons2021}.
In the bare $\mathrm{HM}_\text{b}$ system, the relevant modes are dominated by out-of-plane ionic motion and therefore carry only a weak in-plane zero-point electric field.
The overlayer introduces an in-plane ionic oscillation in the same frequency range, resulting in the polariton modes of the full heterostructure acquiring an increased in-plane component and enhanced zero-point fluctuations in this direction.
As a result, the effect of adding the overlayer is not a uniform change of all vacuum-field components, but a selective enhancement of $E_x^{\rm vac}$ that can extend far beyond the overlayer thickness.

Motivated by the experiment of Ref.~\cite{kerenCavityalteredSuperconductivity2026a}, we then examine how this vacuum field control affects superconductivity.
We consider a two-dimensional superconducting (SC) layer embedded in the first hyperbolic material, and covered by the second.
The superconductor is treated within conventional Eliashberg theory \cite{bardeenMicroscopicTheorySuperconductivity1957,migdal1958interaction,eliashberg1960interactions}.
Within this model, the overlayer remotely modifies the vacuum fluctuations entering the pairing problem and thereby changes the superconducting critical temperature $T_c$.
The response is frequency selective and can persist over distances much larger than the overlayer thickness, capturing two qualitative features observed in the experiment.
Our results identify hyperbolic heterostructures as a platform for making vacuum fluctuations an active design element, enabling long-range control over light--matter couplings and material properties in vdW quantum materials.

\textit{Engineering vacuum fluctuations in hyperbolic materials.---}
Hyperbolic materials (HMs) are a special class of optically anisotropic media.
In a uniaxial medium, the dispersion relation of the transverse magnetic (TM) mode (the extraordinary wave) can be obtained from Maxwell's equations as
\begin{eqnarray}
\frac{k_x^2 + k_y^2}{\epsilon_z(\omega)} + \frac{k_z^2}{\epsilon_x(\omega)} = \frac{\omega^2}{c^2},
\label{eq:TM_dispersion}
\end{eqnarray}
where $\epsilon_x(\omega)$ and $\epsilon_z(\omega)$ are the macroscopic in-plane and out-of-plane diagonal components of the dielectric tensor, respectively, $k_i$ are the components of the wavevector, $\omega$ is the angular frequency, and $c$ is the speed of light in vacuum.
Without loss of generality, we set $k_y=0$ in the following discussion.
When $\epsilon_x \epsilon_z < 0$, the TM mode becomes hyperbolic, and the optical properties of the material are qualitatively different \cite{poddubnyHyperbolicMetamaterials2013,guoHyperbolicMetamaterialsDispersion2020}.
In particular, the wavevector can remain real inside the material even far beyond the diffraction limit, $k_x, k_z \gg \omega/c$.
Equivalently, the isofrequency contour changes from a closed ellipse in an ordinary medium to an open hyperbola, allowing large in-plane momenta inside the HM rather than being restricted to a finite momentum shell~\cite{smithElectromagneticWavePropagation2003,poddubnyHyperbolicMetamaterials2013,guoHyperbolicMetamaterialsDispersion2020}.
As a result, an HM can support multiple oscillating modes inside it, which are evanescent outside.
In this sense, an HM slab confines electric fields on a sub-wavelength scale, in a similar manner to how a Fabry--P\'erot cavity confines free-space optical photons.

There is, however, an important difference for subwavelength light.
In this regime, the energy stored in the magnetic field is much smaller than that stored in the electric field \cite{khurginHowDealLoss2015}.
As a result, the usual picture of an electromagnetic wave, in which the energy oscillates back and forth between electric and magnetic fields, no longer applies.
Instead, the energy oscillates between the electric field and material degrees of freedom, such as plasmons, phonons, or excitons.
This is why these modes are referred to as polaritons 
\cite{basovPolaritonsVanWaals2016,lowPolaritonsLayeredTwodimensional2017a,basovPolaritonPanorama2020,basovPolaritonicQuantumMatter2025}. For hyperbolic polaritons, since the material response is highly directional, their field is also directional.
For example, if $\epsilon_x>0$ and $\epsilon_z<0$, only the material excitation polarized along the $z$ direction is resonant.
Consequently, the in-plane ($x$) component of the confined field is expected to be weak, especially in the large momentum limit.

\begin{figure}
  \centering
  \includegraphics[width=0.98\linewidth]{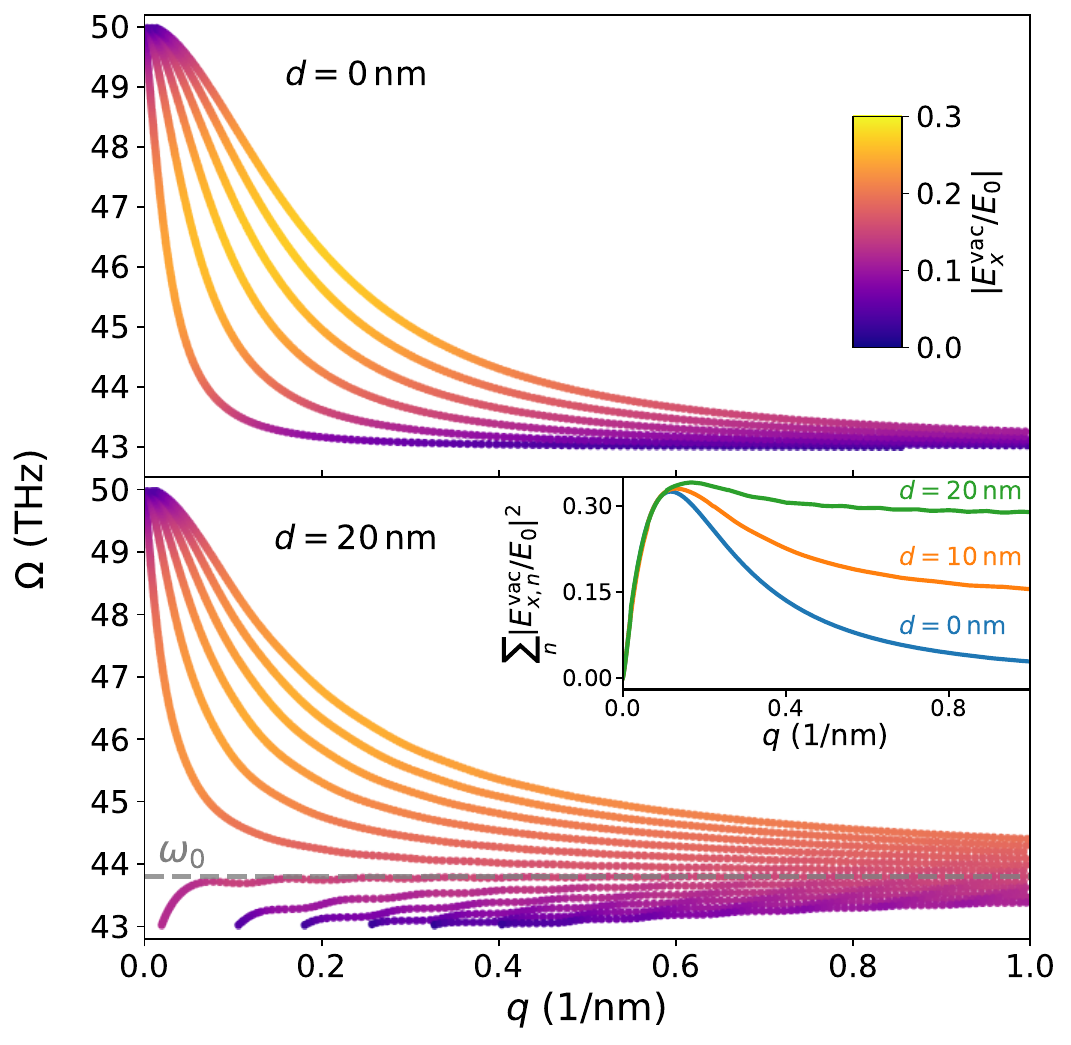}
  \caption{
  \label{fig:dipersion_Ex}
  {\bf Dispersion and in-plane vacuum electric field for the lowest six hyperbolic phonon-polariton branches.}
  The color denotes $|E_x^{\rm vac}/E_0|$ inside the $\mathrm{HM}_{\rm b}$ slab, where $E_x^{\rm vac} = \sqrt{\langle E_x^2\rangle}_{\rm vac}$.
  Top: A bare $\mathrm{HM}_{\rm b}$ slab of thickness $l=100~\mathrm{nm}$ with $\epsilon_x>0$ and $\epsilon_z<0$, where the out-of-plane TO and LO phonon frequencies are $\Omega_{z,\rm TO}^{b}=43~\mathrm{THz}$ and $\Omega_{z,\rm LO}^{b}=50~\mathrm{THz}$.
  Bottom: The same $\mathrm{HM}_{\rm b}$ slab covered by a $\mathrm{HM}_{\rm t}$ overlayer of thickness $d=20~\mathrm{nm}$.
  Hybridization with the overlayer modes shifts the high-momentum components toward $\omega_0$, the frequency set by phase cancellation across the two layers, and enhances the in-plane vacuum field at large $q$.
  The inset shows the branch-summed fluctuations $\sum_n |E_{x,n}^{\rm vac}/E_0|^2$ as a function of $q = |{\bf q}|$ for thicknesses $d = 0$, $10$, and $20~\mathrm{nm}$ of the overlayer, with the $\mathrm{HM}_{\rm b}$ thickness fixed at $l=100~\mathrm{nm}$.}
\end{figure}

To make the directionality of the vacuum field explicit, we first introduce a minimal model for a hyperbolic material, denoted $\rm \mathrm{HM}_b$ (the subscript ``$\text{b}$'' is for bottom, see Fig.~\ref{fig:setup}).
In this model, the hyperbolic response is generated by a finite-frequency matter excitation polarized along a single spatial direction.
As a simple realization, we take this excitation to be an out-of-plane optical phonon, while the in-plane dielectric response is assumed to be non-resonant in this frequency range.
The dielectric function is therefore modeled as
\begin{eqnarray}
    \epsilon_z^b(\omega)
    =
    \epsilon_{z,\infty}^b
    \frac{(\Omega_{z,\rm LO}^b)^2-\omega^2}
    {(\Omega_{z,\rm TO}^b)^2-\omega^2},
    \qquad
    \epsilon_x^b(\omega)
    =
    \epsilon_{x,\infty}^b .
    \label{eq:dielectric_function_1}
\end{eqnarray}
Here we take $\Omega_{z,\rm TO}^b=43~\mathrm{THz}$, $\Omega_{z,\rm LO}^b=50~\mathrm{THz}$, and $\epsilon_{z,\infty}^b=\epsilon_{x,\infty}^b=2$ for the calculations below.
Between the transverse optical (TO) and longitudinal optical (LO) phonon frequencies, $\epsilon_z^b(\omega)<0$ while $\epsilon_x^b(\omega)>0$.
This so-called Reststrahlen band is therefore the hyperbolic regime of $\rm \mathrm{HM}_\text{b}$, and the corresponding hybrid light--phonon modes are hyperbolic phonon polaritons (HPPs) \cite{daiSubdiffractionalFocusingGuiding2015a,caldwellSubdiffractionalVolumeconfinedPolaritons2014,daiSubdiffractionalFocusingGuiding2015a,shiAmplitudePhaseResolvedNanospectral2015,yoxallDirectObservationUltraslow2015,kurmanSpatiotemporalImaging2D2021,guoHyperbolicPhononpolaritonElectroluminescence2025}.
The eigenmodes of a finite $\rm \mathrm{HM}_\text{b}$ slab can be obtained from Maxwell's equations, as described in Ref.~\cite{SM}, and are shown as curves in the top panel of Fig.~\ref{fig:dipersion_Ex}.
The different branches correspond to discrete mode indices $n$, which arise from the finite slab thickness.

Notice that we neglected the damping term in the Lorentzian model.
This simplifies the quantization procedure and does not change our main results, as long as the losses are small, which is usually the case for phonon-based materials \cite{khurginHowDealLoss2015,gilesUltralowlossPolaritonsIsotopically2018,niLongLivedPhononPolaritons2021}.
With this simplification, we can quantize the electric field in the analytical form \cite{SM}
\begin{eqnarray}
\hat{\boldsymbol{E}}(\boldsymbol{r})=\sum_m \sqrt{\frac{\hbar \omega_m}{2\epsilon_0}}[\boldsymbol{F}_m(\boldsymbol{r})\hat{\gamma}_m +\boldsymbol{F}^*_m(\boldsymbol{r})\hat{\gamma}^\dagger_m],
\label{eq:quantized_E}
\end{eqnarray}
where $\boldsymbol{F}_m$ denotes the electric field mode function of the $m$-th mode with frequency $\omega_m$, and ${\hat{\gamma}}^\dagger_m$ ($\hat{\gamma}_m$) is the corresponding polariton creation (annihilation) operator.
Here, $m$ includes both the in-plane momentum $\mathbf{q}$ and the branch index $n$.

The magnitude of the vacuum fluctuations of the in-plane electric field $E_x^{\rm vac} = \sqrt{\langle E_x^2\rangle}_{\rm vac}$ in an $\rm \mathrm{HM}_\text{b}$ slab of thickness $l=100~\mathrm{nm}$ is illustrated by the color scale in the top panel of Fig.~\ref{fig:dipersion_Ex}.
The field amplitude is normalized to \(E_0=\sqrt{\hbar\omega / (2\epsilon_0 l A_{\text{eff}})} \), where $A_{\text{eff}}$ is the effective in-plane mode area.
We note that the in-plane vacuum field becomes very weak at large in-plane momentum $q = |{\bf q}|$, because of the lack of matter support.
As we will discuss later, this is unfavorable for coupling to matter excitations with in-plane polarization, which can usually access a broad momentum phase space.

The polariton dispersion can be engineered by coupling a hyperbolic material to a neighboring layer~\cite{huTopologicalPolaritonsPhotonic2020,chenVanWaalsIsotope2023}.
We now show how such a heterostructure can also control these vacuum fluctuations by placing a second hyperbolic material, denoted $\mathrm{HM}_\text{t}$, with the opposite type of hyperbolicity on top of the original slab, as illustrated in Fig.~\ref{fig:setup}.
If the hyperbolic frequency ranges of the two materials overlap, the electric field can propagate across the interface, and the vacuum fluctuations of the entire structure are significantly modified.
For the calculations below, we take $\rm \mathrm{HM}_\text{t}$ to be hBN, with dielectric function
\begin{eqnarray}
    \epsilon_i^t(\omega)=\epsilon_{i,\infty}^t\frac{(\Omega^t_{i,\rm LO})^2-\omega^2}{(\Omega^t_{i,\rm TO})^2-\omega^2},
    \label{eq:dielectric_function_2}
\end{eqnarray}
where $i\in\{ x,z \}$, the dielectric constants are $\epsilon_{x,\infty}^t = 4.9$ and $\epsilon^t_{z,\infty} = 2.95$, and the phonon frequencies are $\Omega_{x,\rm TO}^t = 40.8$~THz, $\Omega_{x,\rm LO}^t = 48.4$~THz, $\Omega_{z,\rm TO}^t = 22.8$~THz and $\Omega_{z,\rm LO}^t = 24.8$~THz \cite{suFundamentalsEmergingOptical2024a}.
Notice that the in-plane Reststrahlen band of hBN, where $\epsilon_x^t(\omega)<0$, overlaps with the out-of-plane Reststrahlen band of the model material $\rm \mathrm{HM}_\text{b}$, where $\epsilon_z^b(\omega)<0$.
In principle, the capping layer need not itself be hyperbolic: a compatible subwavelength mode, such as a frequency-matched plasmonic surface mode, can also hybridize with the HPPs of $\mathrm{HM}_b$ and modify the vacuum field throughout $\mathrm{HM}_b$.

In the bottom panel of Fig.~\ref{fig:dipersion_Ex}, we show the dispersion and the in-plane component of the vacuum fluctuations inside the original $\rm \mathrm{HM}_\text{b}$ slab of thickness $l=100~\mathrm{nm}$, with an additional $\rm \mathrm{HM}_\text{t}$ slab of thickness $d=20~\mathrm{nm}$ placed on top.
First, we see that the dispersion is modified due to the hybridization of the HPP modes in the two HMs, which shifts the frequencies of the high-momentum modes from the edge of the Reststrahlen band toward its center.
The new frequency $\omega_0$ of these high-momentum modes depends on the ratio between $d$ and $l$, and can be estimated as $\sqrt{-\epsilon^b_x(\omega_0)/\epsilon_z^b(\omega_0)}l \approx \sqrt{-\epsilon^t_x(\omega_0)/\epsilon_z^t(\omega_0)}d$, which corresponds to the condition that the phase accumulated through the two layers cancels.
This feature in the dispersion has already been observed and discussed in recent experiments \cite{sternbachNegativeRefractionHyperbolic2023}.
The corresponding HPP density of states is reconstructed accordingly, with its dominant peak shifting from the bare TO edge toward $\omega_0$, as shown in Ref.~\cite{SM}.
More importantly, when we compute the vacuum fluctuation of the in-plane electric field, we find that $E^{\rm vac}_x$ becomes much stronger at large $q$ when the capping layer is present.
Intuitively, this is because the in-plane phonon inside $\rm \mathrm{HM}_\text{t}$ now provides support for these high-momentum modes.

\begin{figure*}[t]
    \centering

    \subfloat[]{%
        \label{fig:SC_setup}
        \includegraphics[width=0.22\textwidth]{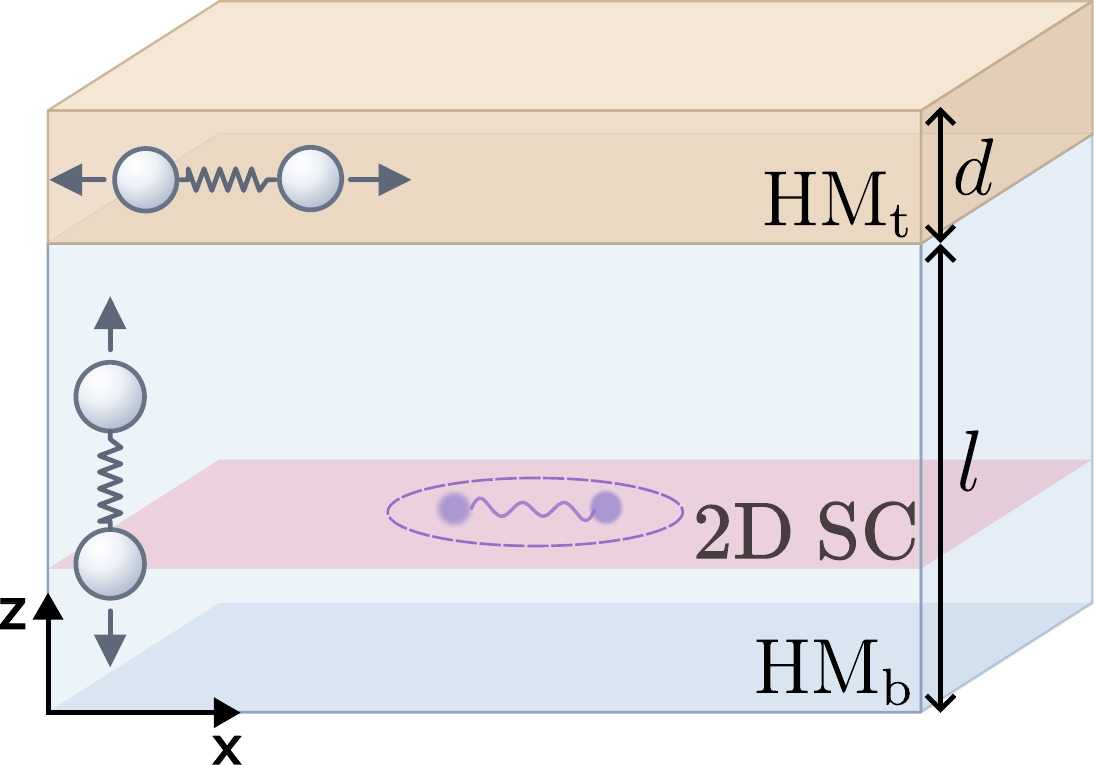}
    }
    \hfill
    \subfloat[]{%
        \label{fig:Tc_d}
        \includegraphics[width=0.35\textwidth]{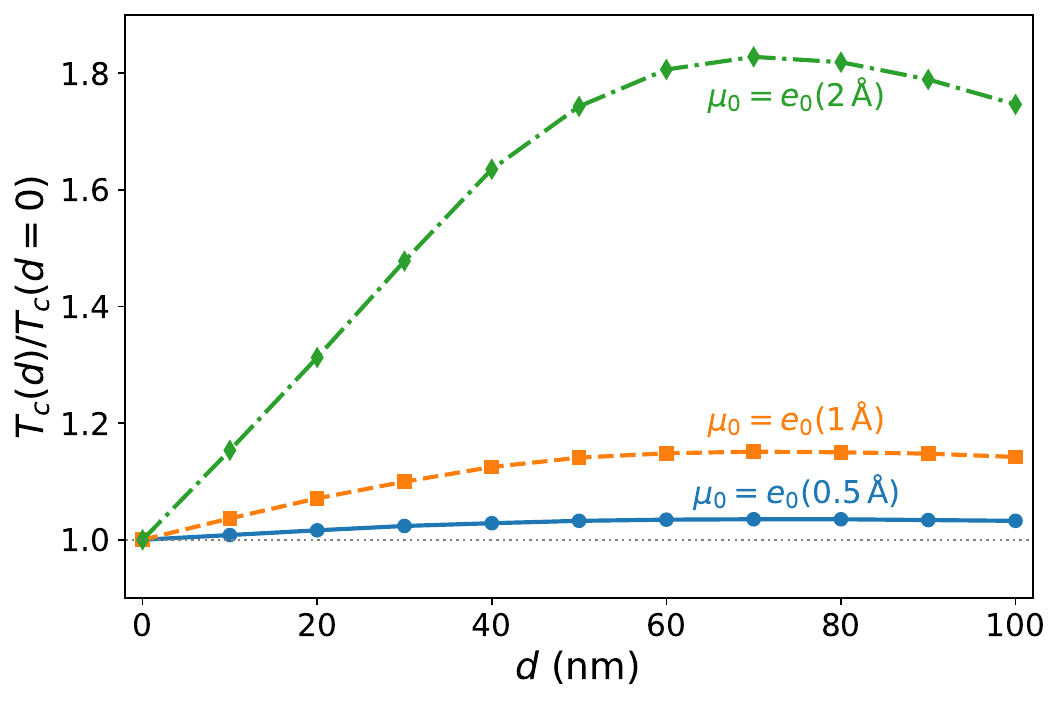}
    }
    \hfill
    \subfloat[]{%
        \label{fig:resonance}
        \includegraphics[width=0.35\textwidth]{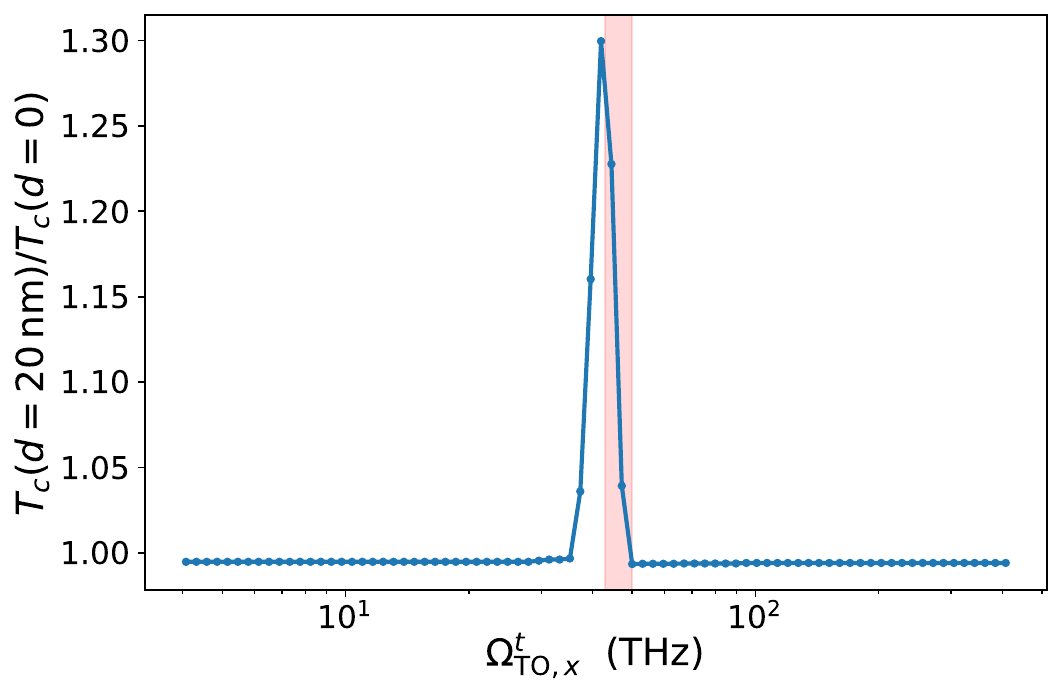}
    }

    \caption{{\bf Hyperbolic control of superconductivity.}
    (a) Schematic of a 2D superconducting layer embedded in a $\mathrm{HM}_\text{b}$ sample of thickness $l$, and covered by an $\mathrm{HM}_\text{t}$ overlayer of thickness $d$.
    (b) Relative change in critical temperature $T_c(d)/T_c(d=0)$, as a function of $d$ for several dipole strengths $\mu_0$.
    The thickness of $\mathrm{HM}_\text{b}$ is fixed to $l=100\,\mathrm{nm}$, with out-of-plane TO and LO phonon frequencies $43\,\mathrm{THz}$ and $50\,\mathrm{THz}$.
    (c) $T_c(d=20\,\mathrm{nm})/T_c(d=0)$ as a function of the in-plane TO frequency of $\mathrm{HM}_\text{t}$, with the LO/TO ratio fixed.
    The red shaded area marks the hyperbolic regime of $\rm HM_b$.
    The strongest modification occurs when the hyperbolic frequency range of $\mathrm{HM}_\text{t}$ overlaps the relevant mode of $\mathrm{HM}_\text{b}$.}
    \label{fig:SC_control}
\end{figure*}

We can further sum the contributions from different branches to obtain the total magnitude of the vacuum fluctuations of the in-plane field as a function of momentum $q$.
This is shown in the inset of the bottom panel of Fig.~\ref{fig:dipersion_Ex}.
We find that for a $\rm \mathrm{HM}_\text{b}$ slab of thickness $l = 100~\mathrm{nm}$, even a $10~\mathrm{nm}$ overlayer can substantially enhance the in-plane vacuum field.
We emphasize that, since the relevant modes extend throughout $\rm HM_b$, the modification can affect embedded excitations far from the interface.
A useful analogy to understand this effect is how changing the properties or position of one mirror in a Fabry-Perot cavity leads to a change in the vacuum fluctuation for the entire cavity.
For thicker $\rm \mathrm{HM}_\text{t}$ slabs, the enhancement becomes even larger.

\textit{Application to a 2D superconductivity inside the hyperbolic medium.---}
We have thus established that the $\mathrm{HM}_\text{t}$ layer can globally modify the in-plane vacuum fluctuation inside $\mathrm{HM}_\text{b}$. 
Therefore, matter excitations in $\mathrm{HM}_\text{b}$ that couple to the in-plane electric field can be controlled remotely by changing the thickness of $\mathrm{HM}_\text{t}$.
One simple example is a localized emitter with a dominant in-plane dipole moment, such as a defect inside $\mathrm{HM}_\text{b}$, whose spontaneous-emission spectrum would be modified by the overlayer.

Here we focus instead on a condensed matter application: a two-dimensional superconducting layer embedded inside $\mathrm{HM}_\text{b}$ and covered by $\mathrm{HM}_\text{t}$, as shown in Fig.~\ref{fig:SC_setup}.
This setup is motivated by the recent experiment on the molecular superconductor $\kappa$-ET~\cite{kerenCavityalteredSuperconductivity2026a}, where an hBN slab placed on top was found to suppress the superfluid density. In $\kappa$-ET, conducting layers are separated by insulating molecular layers and host an out-of-plane stretching mode around $43~\mathrm{THz}$~\cite{mcguireIncoherentInterplaneConductivity2001,buzziPhaseDiagramLightInduced2021}, making it an effectively 2D superconductor with hyperbolic optical properties.

For simplicity, we model the superconducting layer as a conventional phonon-mediated superconductor using Eliashberg theory~\cite{bardeenMicroscopicTheorySuperconductivity1957,migdal1958interaction,eliashberg1960interactions,marsiglioEliashbergTheoryShort2020}. 
The calculation is therefore not intended as a quantitative description of $\kappa$-ET, which is not a conventional BCS electron-phonon superconductor, but rather as a minimal model inspired by the experimental geometry. 
We take the bosonic mode responsible for the pairing to be a single dispersionless (Einstein) phonon of frequency $\Omega_{\rm sc}$, which hybridizes with the HPP modes of the surrounding structure. 
The hybridization redistributes the original phonon glue among the polaritonic modes and thereby changes the effective pairing spectrum entering Eliashberg theory. We describe this hybridization by \begin{eqnarray}
H_{\rm ph-HPP} &=& \hbar \Omega_{\rm sc} b^\dagger b + \sum_m \hbar \omega_m \gamma_m^\dagger \gamma_m \nonumber \\ 
&&+ \sum_m \hbar g_m \left(\gamma_m^\dagger+\gamma_m\right) \left(b^\dagger+b\right), 
\label{eq:Hamiltonian} 
\end{eqnarray} 
where $b^\dagger$ creates a superconducting phonon and $\gamma_m^\dagger$ creates an HPP of frequency $\omega_m$. 
The index $m$ includes both the in-plane momentum ${\bf q}$ and the branch index $n$. 
In the numerical calculations we keep a finite number $n_c$ of HPP branches to avoid divergences, and use $n_c = 20$ as an illustrative cutoff with other choices shown in Ref.~\cite{SM}.

We assume a linear dipole coupling between the superconducting phonon and the in-plane HPP vacuum field, $\hbar g_m=\mu_x E^{\rm vac}_{x,m}$, where $E^{\rm vac}_{x,m}$ is the field shown in Fig.~\ref{fig:dipersion_Ex}. 
The collective phonon dipole is estimated as $\mu_x\approx\mu_0\sqrt{N}=\mu_0\sqrt{n A_{\rm eff}}$, where $\mu_0$ is the dipole moment of one localized phonon excitation, $N$ is the number of coherently participating dipoles, $n$ is their density per area, and $A_{\rm eff}$ is the effective in-plane mode area. 
The factor $A_{\rm eff}$ cancels the corresponding area normalization of $E_x^{\rm vac}$. 
For the estimates below, we take $\mu_0$ to be of order $e_0\times0.1~\mathrm{nm}$ and $n=1/(1~\mathrm{nm})^2$, consistent with typical dimensions of a crystalline unit cell.

After diagonalizing Eq.~\eqref{eq:Hamiltonian}, we use the resulting hybridized boson spectrum as the pairing glue in the Eliashberg equations \cite{SM}. 
For the electronic part, we choose parameters appropriate for a simple weak-coupling BCS reference system such as Al~\cite{morelCalculationSuperconductingState1962,savrasovElectronphononInteractionsRelated1996}: $\lambda_0=0.4$, $\mu^*=0.1$, and $\Omega_{\rm sc}=6~\mathrm{THz}$. 
These parameters give $T_c \approx 1.5$~K before coupling to the HPP modes. Since $\Omega_{\rm sc}$ is much smaller than the HPP frequencies, the main effect of the hybridization is an effective shift of the superconducting phonon spectrum.
In the Eliashberg calculation, we average the pairing kernel over a circular Fermi surface with $2k_F=1~\mathrm{nm}^{-1}$.
The second layer $\rm \mathrm{HM}_\text{t}$ enhances the in-plane vacuum fluctuations for large in-plane momentum.
It therefore enhances the coupling between the HPP vacuum field and the 2D SC, which in turn modifies the effective pairing interaction. 

Fixing the $\mathrm{HM}_\text{b}$ thickness to $l=100~\mathrm{nm}$, we vary the overlayer thickness $d$ and compute $T_c(d)/T_c(d=0)$, as shown in Fig.~\ref{fig:Tc_d}.
The ratio is larger than unity in this model, indicating an enhancement of superconductivity. 
The enhancement initially grows with overlayer thickness and develops a broad maximum at larger $d$.
The effect is generally stronger for larger dipole moments and varies when more HPP modes are included \cite{SM}.
Physically, the 2D superconductor couples efficiently to in-plane fields, whereas the bare $\mathrm{HM}_\text{b}$ phonon resonance is polarized out of plane and therefore couples only weakly to the superconducting layer.
The $\mathrm{HM}_\text{t}$ overlayer supplies the missing in-plane excitation, which is carried through $\mathrm{HM}_\text{b}$ by propagating hyperbolic modes and reaches the embedded superconducting layer.

Although this minimal model is not intended as a quantitative description of the $\kappa$-ET experiment, and gives an enhancement rather than the observed suppression of superconductivity, it nevertheless captures two qualitative features suggested by the data.
The first feature is frequency selectivity.
In Fig.~\ref{fig:resonance}, we rescale the TO and LO phonon frequencies of $\mathrm{HM}_\text{t}$ while keeping their ratio fixed (we set $n_c=5$ and $\mu_0=e_0(2\rm \AA)$ for this illustration), and find that $T_c$ changes appreciably only when the hyperbolic band of $\mathrm{HM}_\text{t}$ overlaps with that of $\mathrm{HM}_\text{b}$.
This is consistent with the experimental observation that superconductivity is modified by hBN, whose Reststrahlen band overlaps with an out-of-plane C=C molecular vibration in $\kappa$-ET, while another in-plane hyperbolic material with a different frequency range does not produce the same effect.
The second feature is the length scale.
Experimentally, a $25~\mathrm{nm}$ hBN slab appears to influence superconductivity over hundreds of nanometers in $\kappa$-ET, far beyond the few-nanometer range expected for an evanescent field at an ordinary dielectric interface.
In our model, the modification is instead carried by propagating hyperbolic modes of $\mathrm{HM}_\text{b}$ and can therefore extend through the slab.
Although the magnitude remains small when the overlayer is too thin, a $10$--$20~\mathrm{nm}$ $\mathrm{HM}_\text{t}$ slab can already produce a noticeable change in $T_c$ for an $\mathrm{HM}_\text{b}$ slab that is an order of magnitude thicker.

\textit{Discussion.---}
Our results show that the vacuum field of a target hyperbolic material can be reshaped by an adjacent hyperbolic control layer.
The role of the control layer is not simply to add another set of polaritonic modes, but to hybridize with the modes of the original slab when the two hyperbolic bands overlap.
Because the two media have different anisotropies, this hybridization modifies the vacuum fluctuations in a directional way.
Since the relevant HPP modes are extended completely through the underlying hyperbolic slab, the modification is not confined to the interface but can affect the whole volume of the heterostructure.
This distinguishes the mechanism from ordinary near-field control and provides a route to tune the vacuum field experienced by embedded matter degrees of freedom.

The superconducting example illustrates one possible consequence of this mechanism.
Even in a minimal Eliashberg model with an Einstein phonon, changing the polaritonic vacuum field modifies the effective pairing glue and hence the critical temperature.
We do not expect the sign or magnitude of this change to be universal, since both depend on the microscopic pairing mechanism and on material-specific properties.
Instead, the general implication is that hyperbolic heterostructures allow the electromagnetic environment to be engineered inside a material, rather than only at a surface or through an external cavity.
Frequency-matched hyperbolic layers can therefore act as long-range control layers for embedded quantum degrees of freedom, including emitters, phonons, excitons, and correlated electronic phases.
These results suggest that optical anisotropy in van der Waals quantum materials is not only a way to control light propagation, but also a knob for shaping the electromagnetic vacuum that governs light--matter interactions inside the material.

\begin{acknowledgments}
We thank Andrew J. Millis, Tatiana A. Webb, Patrick A. Lee, Jingkai Quan, Takuya Okugawa, and Na Wu for fruitful discussions.
This work was supported by the European Research Council (ERC-2024-SyG- 101167294 ; UnMySt), the Cluster of Excellence Advanced Imaging of Matter (AIM), Grupos Consolidados y Alto Rendimiento UPV/EHU, Gobierno Vasco (IT1453-22) and SFB925,  
We acknowledge support from the Max Planck-New York City Center for
Non-Equilibrium Quantum Phenomena. The Flatiron Institute is a division of the Simons Foundation.
DMK acknowledges funding from the Deutsche Forschungsgemeinschaft (DFG, German Research Foundation) - 531215165 (Research Unit ``OPTIMAL'').
\end{acknowledgments}

\bibliographystyle{apsrev4-2}
\bibliography{reference}

\end{document}